# Optical Properties of Fully Inorganic, Spacer-Free Core/Gradient-Shell CdSe/CdZnS Nanocrystals at the Ensemble and Single-Nanocrystal Levels[†]

Justine Baronnier[‡], Benoit Mahler[‡], Christophe Dujardin[‡], Florian Kulzer[‡], and Julien Houel[‡*]

June 23, 2021

We report the synthesis and optical characterization of fully inorganic gradient-shell CdSe/CdZnS nanocrystals (NCs) with high luminescence quantum yield (QY, 50 %), which were obtained by replacing native oleic-acid (OA) ligands with halide ions ($Br^-$ and $Cl^-$). Absorption, photoluminescence excitation (PLE) and photoluminescence (PL) spectra in solution were unaffected by the ligand-exchange procedure. The halide-capped NCs were stable in solution for several weeks without modification of their PL spectra; once deposited as unprotected thin films and exposed to air, however, they did show signs of aging which we attribute to increasing heterogeneity of (effective) NC size. Time-resolved PL measurements point to the existence of four distinct emissive states, which we attribute to neutral, singly-charged and multi-excitonic entities. We found that the relative contribution of these four components to the overall PL decay is modified by the OA-to-halide ligand exchange, while the excited-state lifetimes themselves, surprisingly, remain largley unaffected. The high PL quantum yield of the halide-capped NCs allowed observation of single particle blinking and photon-antibunching; one surprising result was that aging processes that occurs during the first few days after deposition on glass seemed to offer a certain increased protection against photobleaching. These results suggest that halide-capped CdSe/CdZnS NCs are promising candidates for incorporation into opto-electronic devices, based on, for example, hybrid perovskite matrices, which require eliminating the steric hindrance and electronic barrier of bulky organic ligands to ensure efficient coupling.

---

# 1 Introduction

Cadmium-selenide-based nanocrystals (CdSe NCs) have become indispensable for numerous applications in the last twenty years [1–8]. This success was built on the remarkable optical properties of these nanoparticles – high quantum yield [9, 10], fast exciton dynamics [11] with narrow emission spectra [12], tunability over the visible spectrum [7] – combined with highly reproducible, industrial-scale synthesis [13]. An important requirement for technological applications is the possibility of incorporating these versatile quantum emitters into fully inorganic [4] or hybrid organic/inorganic devices [14]. A particularly attractive class of materials of the latter kind are hybrid perovskites, whose electrical and optical properties can be tuned by changing their composition, which made it possible to engineer record-breaking photovoltaic cells [15] and other light-harvesting devices [16–18]. Integrating semiconductor quantum dots into these matrices obviously has great potential, provided that the surface chemistry of the NCs con be controlled to ensure compatibility, electronic coupling and conservation of the favorable properties of both guest and host. Functionalizing the surface of NCs has in fact been a crucial issue ever since the first synthesis strategies were developed [13]; as core-only NCs consist of only a few hundred atoms, most of them located at or close to the surface [13, 19], which therefore strongly influences the optical properties of such nanoparticles [20]; moreover, modifying the surface of CdSe nanocrystals has recently ben shown to be of great interest for photocatalysis applications[21, 22]. When core/shell NCs became established [23], ligand tuning was used to control fluorescence intermittency [24–27] and more recently ligand-exchange protocols were developed to optimize the coupling of NCs to a host material [14] or to improve their contribution to conductivity in thin films [28]. Ning *et al.* succeeded in incorporating lead-sulfide NCs in a hybrid perovskite matrix [14], which required changing the passivation layer of the NC surface from organic ligands to halide ions to achieve compatibility with the perovskite host. Two drawbacks of PbS NCs, however, are their low luminescence quantum yield (QY) and their emission in the near infrared. In view of potential applications for high-efficiency devices operating in the visible spectral range, it would therefore be interesting to incorporate high-QY CdSe-based NCs in halide perovskite materials. A number of strategies for surface-functionalization of CdSe NCs with halide ligands have already been investigated [20, 28–34]. Purcell-Milton *et al.* [33] have recently published a study of the evolution of the quantum yield of CdSe-based NCs that occurs when organic ligands such as oleic acid (OA) are replaced with halide ions ($Br^-$, $Cl^-$ and $I^-$), which resulted in an increase in some cases (CdSe core-only NCs, type II CdS/CdSe) and a rather low influence for CdSe/CdS; in the case of CdSe/ZnS, however, a strong reduction of the QY was observed. In the present article, we focus on CdSe/CdZnS core/gradient-shell NCs [35–38] to take advantage of their hybrid band structure, which offers a combination of the advantages of, on one hand, quasi-type-II large-shell CdSe/CdS NCs, which are known be highly photostable [39], and of, on the other hand, type-I CdSe/ZnS NCs, whose electrons are well-confined in the CdSe core, which leads to short emission lifetimes. [40] Furthermore, as CdSe/CdZnS NCs are already well established for various applications, [41–43] it is tantalizing to envision their incorporation in perovskite devices. While there have been previous reports on the optical properties of surface-modified CdSe-CdS/ZnS [44, 45] and CdSe/CdZnS [46] NCs, these studies so far did not include, to the best of our knowledge, the case of halide ligands. Here we report the synthesis of robust, highly luminescent, type I core/gradient-shell CdSe/CdZnS NCs whose surface is passivated with $Br^-$ and $Cl^-$ ions, and we present an extensive study of their optical properties compared to the OA-capped precursors, at the ensemble level *via* steady-state spectroscopy and time-resolved emission in solution and thin films, as well as for individual nanoparticles in terms of their luminescence blinking dynamics and single-photon emission.

## 2 Experimental Section

### 2.1 Nanocrystal Synthesis

#### 2.1.1 Chemicals

1-Octadecene (ODE, 90 %) and oleic acid (OA, 90 %) were purchased from Alfa Aesar. Myristic acid (99 %) was obtained from ACROS. 1-dodecanethiol (DDT, 98 %), trioctylphosphine (TOP, 97 %), and zinc oxide (ZnO, 99.99 %) were obtained from Sigma-Aldrich. Selenium powder (99.99 %) Sulfur (99.999 %), and cadmium oxide (CdO, 99.99 %) were purchased from Strem. All chemicals were used without further purification.

#### 2.1.2 Preparation of Precursors

TOPSe 2 M: 20 mmol of selenium powder were added to 10 mL of TOP in a glovebox and magnetically stirred at room temperature for 24h. TOPS 2 M: 20 mmol of sulfur and 10mL of TOP were introduced in a 50 mL three-neck flask and heated to 100 °C for 1 h under argon. The resulting precursor solution was stored in an inert-atmosphere glovebox. $Zn(OA)_2$ 0.5 M: 50 mmol of Zinc acetate, 50 mL oleic acid and 50 mL ODE were introduced in a three-neck flask. The mixture was heated to 250 °C under stirring for 1 h until it became transparent. The temperature was then decreased to 110 °C and the solution was degassed for 2 h under vacuum to remove the produced water. After preparation the precursor was stored in an inert-atmosphere glovebox. $Cd(OA)_2$ 0.5 M: 50 mmol of CdO, 50 mL oleic acid and 50 mL ODE were introduced in a three-neck flask. The mixture was heated to 300 °C under stirring until it became transparent. The temperature was then decreased to 100 °C and the solution was degassed under vacuum to remove the produced water. After preparation the precursor was stored in an inert-atmosphere glovebox.

#### 2.1.3 Oleate-Capped Core/Shell Nanocrystals

The preparation of oleate-capped CdSe/CdZnS core/shell nanocrystals was adapted from the one-pot synthesis described by Lim *et al.* [41] Briefly, 15 mL octadecene (ODE), 1 mmol of cadmium oxide (CdO) and 3 mmol of myristic acid (MA) were introduced in a three-neck flask and pumped for 10 minutes at 60 °C. The solution was then heated up to 300 °C under Ar atmosphere. At transparency, 0.25 mL TOPSe at 2 M were introduced, followed by a three-minute annealing period, which produced CdSe cores of 5 nm diameter. The initial layer of the shell was then grown by injecting a mixture of 3 mL zinc oleate $[Zn(OA)_2]$ at a concentration of 0.5 M and 1 mmol of dodecanethiol (DDT), followed by annealing for 30 minutes at 300 °C. Subsequent shell layers were added by simultaneously injecting 2 mL of cadmium oleate $[Cd(OA)_2]$ at a concentration of 0.5 M, 4 mL of $Zn(OA)_2$ at 0.5 M, and 1.5 mL TOPS at 2 M, followed by 10 minutes of annealing; this last step was repeated four times. After cooling down, the obtained nanocrystals were purified through precipitation/redispersion using ethanol and toluene, respectively, and stored in 10 mL hexane. The above protocol yielded CdSe/CdZnS NCs capped with oleic-acid ligands, with a diameter of $(12.6 \pm 1.4)$ nm and a measured quantum yield of 70 % at room temperature.

#### 2.1.4 Ligand-Exchange Protocol

To replace the oleate surface-ligands by halide ions, we developed a procedure based on the work of Ref. 30: 60 µL of as-synthesized oleate-capped nanocrystals were mixed with 2 mL of hexane, then 2 mL of a solution of methylamonium bromide (MABr) at 0.1 M in dimethylformamide (DMF) and 5 mg of Cadmium bromide (CdBr) were added. The resulting biphasic solution was magnetically stirred for 2 hours at room temperature, leading to aggregation of the nanocrystals at the hexane/DMF interface. After precipitation in isopropanol and centrifugation, the precipitated NCs were taken up in 3 mL of Hexane, to which 1 mL of methylamonium chloride (MACl) in DMF (0.03 M) was added. The resulting solution was sonicated for 30 minutes at 50 °C. The above protocol resulted in CdSe/CdZnS NCs capped with $Br^-$ and $Cl^-$ ligands, which exhibited a quantum yield of 50 % at room temperature.

### 2.2 Optical Spectroscopy

Optical spectroscopy was performed with a commercial spectrofluorometer (Edinburgh Instruments FS 5) equipped with a double spectrometer and a Xenon white-light lamp. Photoluminescence measurements were carried out with 405 nm excitation wavelength, 1 nm excitation width and 1 nm detection resolution. All quantum yields given in the article were measured with the three-point method [47] relative to a fluorescent standard under identical conditions. Fourier-transform infrared (FTIR) spectroscopy was carried out with a commercial infrared micro-spectrometer (Perkin Elmer FT-IR GX-Auto-image) in total-internal-reflection geometry. Lifetime measurements on QDs in solution and deposited as thin-films on glass were performed using a home-built microscope coupled to a pulsed source at 1.95 MHz, with an emission wavemength of 450 nm and 10 nm emission width. To minimise photo-induced damage to the thin-films, measurements were performed while scanning the sample over 50x50 $\mu m^2$ at a rate of 10 $\mu$m/s.

### 2.3 Single-QD Measurements

Single-particle measurements were carried out with a home-built confocal microscope described previously [48]. The excitation laser was focused onto the sample (bare QDs on a glass substrate) with an oil-immersion objective (Olympus 60×, NA 1.35), which resulted in a spatial resolution of 150 nm. The photoluminescence of individual NCs was collected through the same objective and sent to an Hanbury Brown and Twiss (HBT) interferometer with two single-photon avalanche photodiodes (SPADs, Perkin Elmer ARQH-50 and Laser Components Count-20). Luminescence timetraces were recorded under continuous-wave (CW) excitation at $\lambda = 561$ nm with an intrinsic resolution of 1 ms, which was post-processed to 10 ms for the analysis of blinking statistics. Photon-coincidence measurements were performed under CW excitation with a laser diode at 405 nm; the overall temporal response of this Hanbury Brown–Twiss (HBT) setup was 1.2 ns.

### 2.4 Analysis of Time-Resolved Photoluminescence Measurements

Photoluminescence decay curves and photon-coincidence histograms were analyzed with a maximum-likelihood technique [49] based on multinomial likelihood functions [50] and models that included convolution with the instrumental response. Photon-coincidence histograms were described with a single-exponential model [51], while photoluminescence decay curves were fitted with models of up to four exponential components; in the latter case, likelihood-ratio tests [49] were employed to quantify the statistical significance of choosing a more complex model with a greater number of free parameters. The uncertainties of the best-fit model parameters were determined by bootstrap resampling [52]. Detailed descriptions of all these analysis techniques are provided in the ESI†.

### 2.5 Analysis of Intensity Timetraces

The analysis of fluorescence intensity timetraces started with compiling a histogram of the observed number of counts per timetrace bin. If this histogram showed a clear "off" level separated from one or more "on" levels by an intermediate dip in the count histogram then the part of the histogram below this dip was fitted with a Poisson distribution to obtain an estimation of the average "off" level. If there was no discernible dip in the histogram then the rising flank of the main peak was used for the Poisson adjustment. Having identified an average "off" number of counts, a threshold for identifying an "on" bin at a given level of statistical significance, typically $p = 0.05$, could be determined from the corresponding cumulative Poisson probability function. The number of timetrace bins reaching or surpassing this threshold was then divided by the total number of timetrace bins to obtain the fraction of time spent in the "on" state. It should be noted that the values determined in this manner are intended as simple heuristic estimates of the "on" state contribution, whose main purpose was to provide a relative measure for comparing different samples, e.g. before and after ligand exchange and/or aging effects. The statistical significance of observed differences in the distributions of this parameters for different samples was quantified with a Kolmogorov–Smirnov (K-S) test.

## 3 Results and Discussion

This presentation and discussion of the optical properties of OA- and halide-capped NCs is organized as follows: First, steady-state spectroscopy is used to document the conservation of the optical properties of the nano-emitters after ligand exchange, as well as to study the stability of the two species in solution and when dispersed on glass. Second, time-resolved photoluminescence measurements serve to elucidate the emission dynamics and the nature of the emitting states, as well as to confirm single-photon emission and the successful detection of individual NCs. Finally, emission timetraces of single NCs are analyzed to compare the blinking behavior, emissivity and photostability of the two NC varieties.

### 3.1 Steady-State Optical Spectroscopy

We present in Fig. 1 a and b the absorbance and photoluminescence excitation (PLE) spectra of the synthesized OA-capped NCs dispersed in hexane (red lines). The first exciton resonance is estimated to correspond to an absorption wavelength of around 620 nm. The overall shapes of the PLE and absorbance spectra as well as their correspondence – the difference in slope below 470 nm, which is attributed to weak reabsorption effects in the PLE spectrum, notwithstanding – agree with what is expected for a core/shell structure [39]; further PLE spectra documenting the growth process of the ZnS shell can be found in the ESI† (Fig. S1). Fig. 1 a and b furthermore show the absorbance and PLE spectra of a solution of halide-capped NCs in DMF (blue lines), which are similar in shape to the corresponding spectra of the OA-capped QDs, indicating that no major modification of the overall band structure of the exciton was induced by the ligand exchange. This conclusion is further supported by the comparison of the photoluminescence (PL) emission spectra of the two species, Fig. 1 c, which are identical in both center wavelength ($\lambda_c = 632$ nm) and full width at half maximum (FWHM, $\Delta\lambda = 30$ nm); furthermore, the FWHM of the emission is consistent with the size dispersion of the OA-capped NC batch as deduced from TEM images of 50 NCs as $(13.0 \pm 2.3)$ nm (see Fig. S2 of ESI† for images). The FTIR spectra shown in Fig. 1 d exhibit the characteristic oleate resonance around $1550\,\text{cm}^{-1}$ only for the OA-capped NCs, while FTIR spectrum of the halide-capped QDs only shows residual DMF absorption bands; the absence of the oleate vibrational band for the halide-capped NCs confirms, in combination with their stability in the 0.015M MACl in DMF solution, that a quantitative replacement of the OA surface ligands by halide ions $Cl^-$ and $Br^-$ has indeed taken place. The conservation of the luminescence quantum yield by the ligand-exchange procedure is illustrated in Fig. 1 e, which compares photographs of the two solutions under both white-light and UV illumination; the measured quantum yields at room temperature were 70 % and 50 %, respectively, for the OA-capped and the halide-capped NCs.

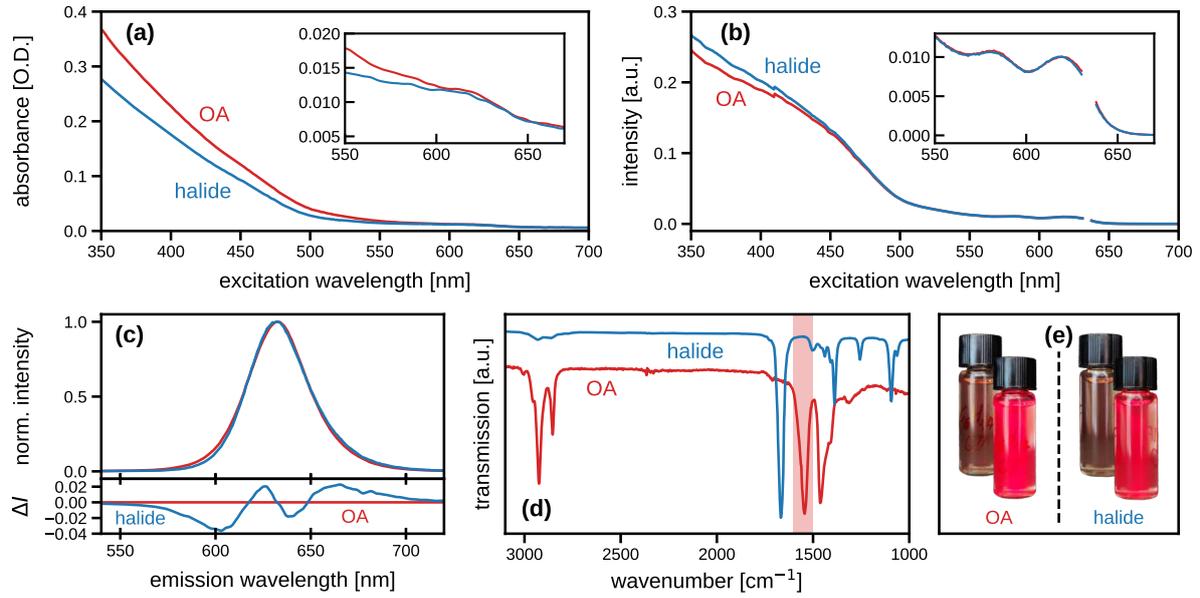

**Figure 1:** Optical properties of CdSe/CdZnS nanocrystals before and after ligand exchange. **(a)** Absorbance spectra of OA-capped NCs in hexane (red) and halide-capped NCs in DMF (blue) **(b)** Photoluminescence-Excitation (PLE) spectra of OA- (red) and halide-capped (blue) NCs in solution. [Breaks in the spectra correspond to the masked detection window, $\lambda_{\text{det}} = (634 \pm 3)\,\text{nm}$; kinks at 410 nm are instrumental artifacts.] **(c)** Photoluminescence (PL) emission spectra of OA- (red) and halide-capped (blue) NCs in solution, recorded on the day of the ligand exchange procedure, $\lambda_{\text{exc}} = 405\,\text{nm}$. The bottom panel shows the intensity difference $\Delta I$ between the two spectra. **(d)** FTIR spectra of OA- (red) and halide-capped (blue) NCs. A vertical stripe marks the characteristic spectral region of carboxylic-bond vibrations around $1550\,\text{cm}^{-1}$. **(e)** Photographs of the solutions of both NC species, taken under white-light (background images) and UV (foreground images) illumination, respectively.

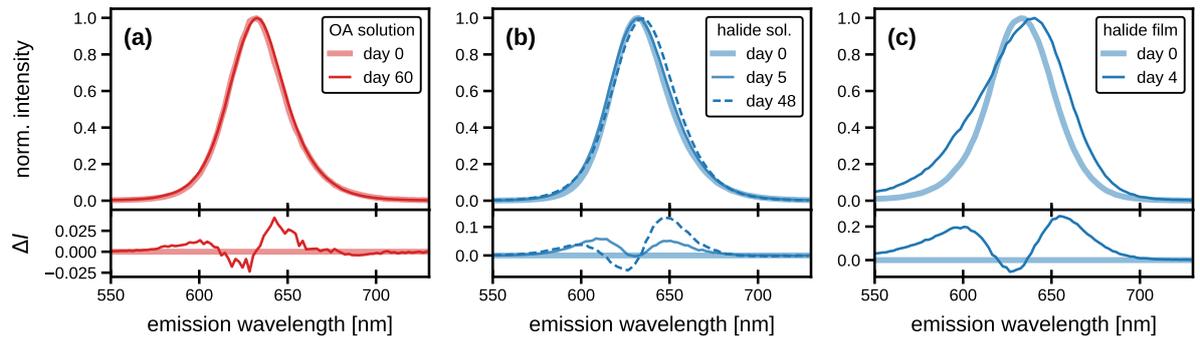

**Figure 2:** Ligand- and environment-dependent stability and aging of NC PL spectra. **(a)** Emission spectra of OA-capped NCs in hexane on the day of synthesis and two months later; the bottom panel shows the intensity difference $\Delta I$ between the spectra. **(b)** Emission spectra of halide-capped NCs in DMF on the day of the ligand exchange, five days later and after 48 days. **(c)** Emission spectra of halide-capped NCs dispersed as a thin film on a glass substrate, on deposition day and four days later.

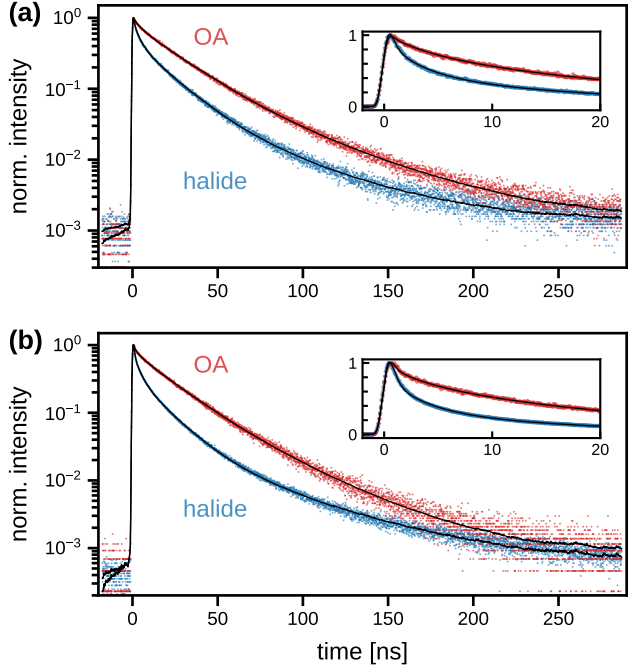

**Figure 3:** Log-scale luminescence decay histograms of OA- (red dots) and halide-capped (blue dots) NCs, fitted with quadri-exponential decay models on constant background (black curves); insets show the the first 20 ns of the decay curves on a linear scale. **(a)** Measurements in solution: OA-capped NCs in hexane, halide-capped NCs in DMF. **(b)** Measurements of thin films obtained by spin-casting the solutions from (a) onto glass substrates without any protective matrix. (The noise on the model curves originates from the measured instrumental response function.)

Figure 2 provides an overview of the evolution of the emission spectra of OA- and halide-capped NCs over time. As can be seen in Fig. 2 a, the PL spectrum of OA-capped NCs in hexane solution is stable over a period of two months. The PL spectrum of halide-capped NCs in DMF, Fig. 2 b, remained unchanged after 5 days in DMF, while after 48 days a redshift of about 2 nm was observed along with a slight asymmetrization of the spectrum. This aging phenomenon of the PL spectrum was drastically accelerated when the halide-capped NCs were spin-cast on a glass slide and left exposed to ambient conditions: After 4 days, we observed a redshift of the emission as large as 5 nm, and an increase of the blue part of the spectrum, leading to pronounced asymmetry of the emission band. Both the broadening and the asymmetrization of the PL spectrum are compatible with the established expression for exciton energies [53] if one assumes a single-QD emission linewidth of 40 meV ($\approx 1.5\,k_\text{B}T$) and an effective confinement radius that is distributed normally around an average of 3.8 nm, with a relative standard deviation that increases from 8 % (day 0) to 14 % (day 4) due to aging. Nevertheless, as we discuss in Section 3.3, the stability of individual QDs dispersed on glass at lower concentrations is much higher than what Fig. 2 c might seem to suggest. As far as emission intensity is concerned, we did observe a two-fold decrease for the halide-capped NCs over the first day after film deposition, but the intensity subsequently remained constant for eight days. (See ESI† for further data and details of the aging model.)

### 3.2 Time-Resolved Photoluminescence Measurements

Figure 3 shows ensemble photoluminescence decay curves of the two NC species in solution and dispersed as a thin film (without any protective matrix) on a glass substrate. All decay curves were maximum-likelihood (ML) fitted with a model featuring four distinct decay components on a constant background and incorporating convolution with the instrumental response function.

**Table 1:** The main fit parameters of the decay models shown in Fig. 3 for OA- and halide-capped NCs in solution and in a concentrated thin film. Listed are the four decay times $\tau_i$ and their respective contributions $f_i$ to the total signal (minus the constant background); uncertainties of the fit parameters were deduced by bootstrap resampling. The excitation intensity used for these measurements, $I_{\text{exc}} \approx 50\,\text{W/cm}^2$, was chosen to be large enough for reliable observation of all four decay components without saturation effects, cf. the data points at $P_{\text{exc}} = 2\,\mu\text{W}$ in Fig. 4.

| | sample | $\tau_1$ [ns] | $f_1$ [%] | $\tau_2$ [ns] | $f_2$ [%] | $\tau_3$ [ns] | $f_3$ [%] | $\tau_4$ [ns] | $f_4$ [%] |
|---|---|---|---|---|---|---|---|---|---|
| sol. | OA | $49.6 \pm 0.8$ | $31 \pm 1$ | $22.3 \pm 0.3$ | $65 \pm 1$ | $3.5 \pm 0.2$ | $3.1 \pm 0.2$ | $0.2 \pm 0.1$ | $1.2 \pm 0.1$ |
| sol. | halide | $47.3 \pm 1.0$ | $24 \pm 1$ | $17.5 \pm 0.2$ | $60 \pm 1$ | $3.3 \pm 0.2$ | $12.4 \pm 0.3$ | $0.3 \pm 0.1$ | $3.6 \pm 0.1$ |
| film | OA | $38.5 \pm 0.2$ | $35 \pm 1$ | $19.4 \pm 0.1$ | $61 \pm 1$ | $2.9 \pm 0.1$ | $2.8 \pm 0.1$ | $0.2 \pm 0.1$ | $1.6 \pm 0.1$ |
| film | halide | $50.6 \pm 0.3$ | $19 \pm 1$ | $15.6 \pm 0.1$ | $56 \pm 1$ | $3.6 \pm 0.1$ | $18.2 \pm 0.1$ | $0.4 \pm 0.1$ | $6.6 \pm 0.1$ |

The goodness-of-fit improvement from adding the fourth component was statistically significant at a high level, $p < 10^{-12}$; a graphical illustration of the goodness-of-fit as a function of the number of exponentials can be found in Figs. S5 and S6 of the ESI† for the two QD decay curves of Fig. 3 a. The ML-analysis results are summarized in contributions $f_i$ to the signal photons, *i.e.*, photons in excess of the time-independent background of each curve. Surprisingly, the lifetimes found for OA- and halide-capped QDs were very similar and did not change much for each species going from solution to thin films. This invariability of the lifetimes makes it seem unlikely that coupling exists for the halide-capped NCs, contrary to what one might have expected for QDs whose small ligands allow for tight packing in a thin-film sample [54]; given that QD dynamics usually depend on their surface states [55–57], a possible explanation for the observed lifetime robustness is a large potential barrier that protects the exciton from environmental influence, leading to a confinement closer to type I than quasi-type II. The relative weights of the decay components, on the other hand, were affected by the OA-to-halide ligand exchange: While component $\tau_2$ remained dominant and accounted for about 60 % of detected signal photons from both QD species, the weight of the longest time $\tau_1$ was reduced in the halide-capped NCs with a concomitant increase (about four-fold) of the strength of the two fastest components $\tau_3$ and $\tau_4$.

The longest time $\tau_1 \approx 40 - 50\,\text{ns}$ lies between the values of the neutral-exciton ($\text{X}_0$) recombination lifetimes of quasi-type-II CdSe/CdS NCs of comparable size (60 ns [39, 58, 59]) and of type-I CdSe/ZnS NCs (20 ns [51, 60]). The power dependence of the number of detected photons in this $\tau_1$ component was linear at low excitation powers, as shown in Fig. 4 a and e, in accordance with the behavior of a single exciton; saturation effects at higher excitation power can be explained with a simple two-level saturation model (see ESI†). We therefore attribute $\tau_1$ to the $\text{X}_0$ recombination lifetime in our OA- and halide-capped CdSe/CdZnS NCs, which both establish a smooth confinement potential that is an intermediate between type I and quasi-type II.

The second-longest lifetimes, $\tau_2$, were a factor of 2.2 and 2.0 smaller than $\tau_1$ for the OA-capped NCs in solution and in the film respectively; the corresponding factors for the halide-capped species were 2.7 and 3.3. As these ratios are close to the factor of 2 expected for the negatively charged exciton [61–63], we tentatively assign $\tau_2$ to the emission of $\text{X}^{1-}$, with a QY comparable to that of $\text{X}_0$ for both QD species, compatible with the measured emission quantum yields (QY) of 70 % (OA) and 50 % (halide). Large NCs have in fact been shown to have preferred emission through the $\text{X}^{1-}$ channel [55–57], consistent with the dominant contribution (60 % of detected photons) of this component in our measured decay curves. The linear power-dependence with saturation of the number of detected photons attributable to $\tau_2$, Fig. 4 b and f, is also in agreement with a single-exciton state such as $\text{X}^{1-}$.

The photons detected for the $\tau_3$ components ($\approx 3-4\,\text{ns}$) also exhibited linear behavior with increasing excitation power followed by saturation, as shown in Fig. 4 c and g, again suggesting single-exciton emission. As it has been surmised previously that the emission lifetime of the positively charged exciton $X^{1+}$ is shorter than that of $X^{1-}$ (our $\tau_2$ candidate) in CdSe/CdS [62], CdSeS/ZnS NCs [40] and CdSe/CdS/ZnS [57], we assign $\tau_3$ to the emission of $X^{1+}$. Finally, the number of photons detected from the fastest component ($\tau_4 < 0.5$ ns) are the only ones whose power dependence is a convex function, see Fig. 4 d and h, implying a mixed linear (single exciton) and quadratic (bi-excitonic) [64, 65] behavior, which leads us to assign $\tau_4$ to a combination of fast emission from Auger-quenched multi-charged single excitons and bi-excitons. It should be noted that $\tau_4$ approaches to our instrumental resolution, meaning that we can interpret its value only as approximate upper boundary; our statistical analysis nevertheless establishes that adding such a short-lived component does in fact improve the data-model agreement significantly ($p < 10^{-12}$).

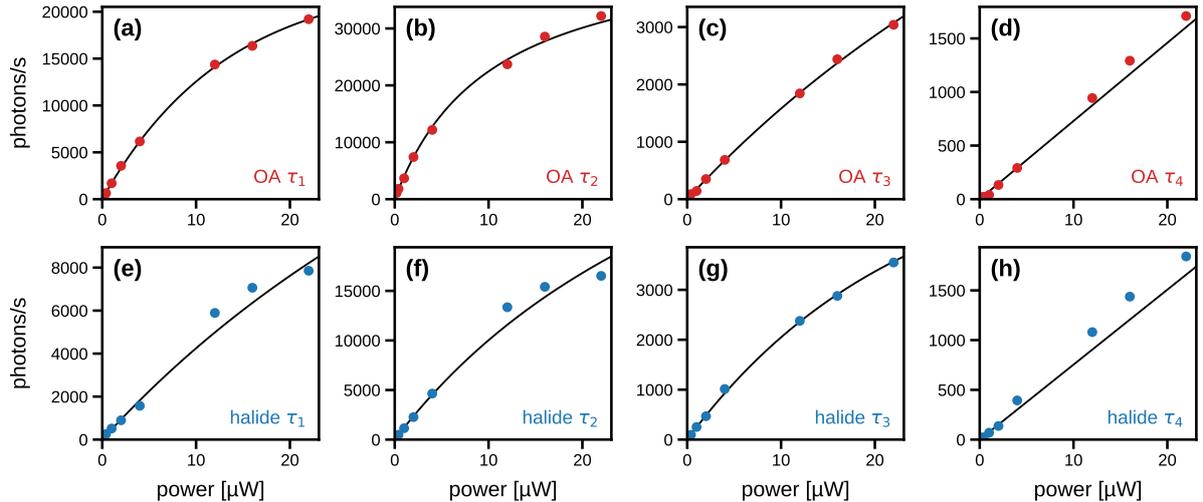

**Figure 4:** The photon detection rates for the four decay components $\tau_i$ as a function of the excitation power (filled circles). Panels **(a)** to **(d)** show this data for OA-capped NCs measured in hexane solution, while panels **(e)** to **(h)** deal with halide-capped QDs in DMF. The solid lines are fits of a linear two-level saturation model detailed in the ESI†. The estimated uncertainties of the measured photon detection rates correspond to error bars smaller than the circles used to represent the data in all cases. The time-independent background in these measurements showed a low-power residual intensity of 80 counts/s, the known dark count rate of our detector, and grew linearly with laser power as expected (data not shown).

The unchanged lifetimes $\tau_1$ to $\tau_4$ indicate that the emitting entities themselves are preserved when switching from an OA-saturated to a halide-rich NC surface, while the emission of charged and multi-excitons is enhanced in halide-capped NCs at the expense of the neutral exciton; to our knowledge, this constitutes the first observation of such a ligand-mediated redistribution of emission probability under preservation of excited-state lifetimes. Obtaining this type of complementary information about the link between preferred emission states of NCs and their surface-treatment could be useful for example in photocatalysis research [21, 22].

## 3.3 Single-QD Blinking

The multi-state emission behavior observed in the photoluminescence decays presented in Section 3.2 corresponds to blinking in the emission of single NCs [19, 66] under continuous excitation, and analyzing these stochastic transitions between states of distinct emissivity – one or more ON or "bright" state(s) on one hand and one or more OFF or "dim" state(s) on the other hand – has proven itself an effective tool to explore differences in surface states of NCs [67–69].

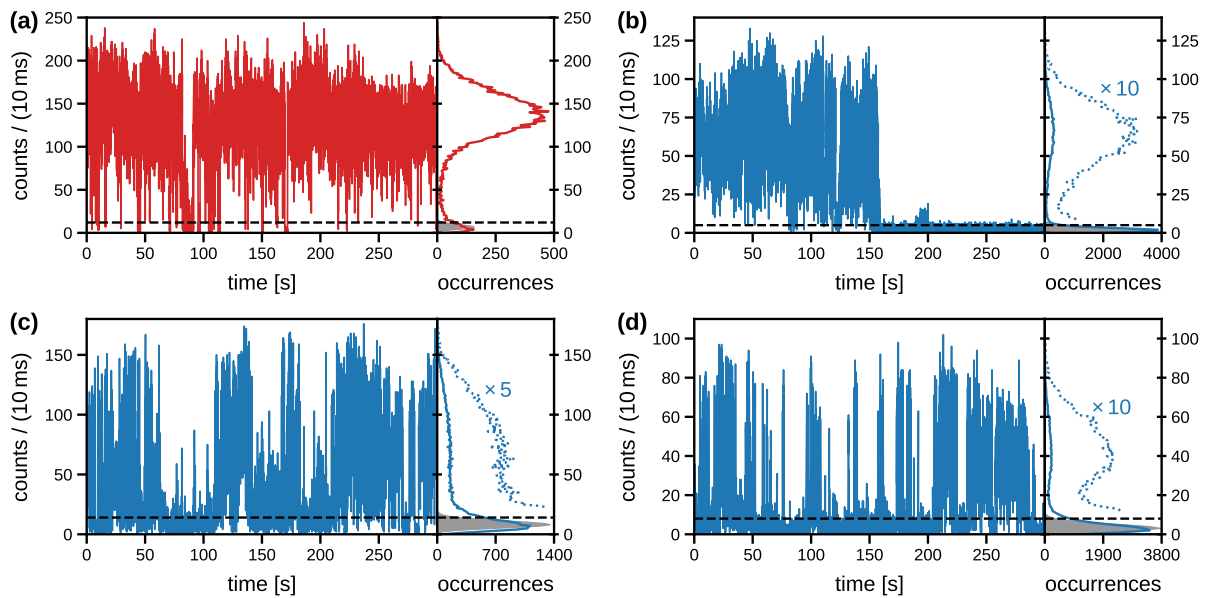

**Figure 5:** Blinking behavior of individual OA- and halide-capped NCs on glass substrates. Main panels show single-QD intensity timetraces binned to 10 ms, while side panels present the corresponding count distributions with transposed $x$ and $y$ axes to allow for visualization of each timetrace/histogram pair with a common ordinate. Solid lines in the side panels are outlines of the count histograms, dotted lines show the subdominant regions of the histograms at the specified factor of magnification along the occurrences axis, and gray areas indicate Poisson fits to the dim-state histogram peak, which served to establish the $p = 0.05$ threshold (dashed horizontal lines) for probabilistically classifying the timetrace bins as "dim" (off) and "not dim" (on). The figure shows examples of **(a)** one OA-capped NC and three halide-capped NCs measured on **(b)** day 0 (deposition day), **(c)** day 1, and **(d)** day 2, respectively.

Even when a single NC remains in a highly emissive state for an extended period, there will be a delay between photon emissions as singly-excited states cannot emit more than one photon at once. This lack of photon coincidences at zero lag time (sub-nanosecond to nanosecond timescale) is in fact a decisive criterion to prove that the detected photons originate from one individual emitter, and we have systematically confirmed the presence of this antibunching phenomenon for the single-QD data that we present (see Fig. S7 of the ESI† for examples). The main interest in of the present section, however, is blinking in the photoluminescence timetraces of individual OA- and halide-capped NCs on the millisecond timescale, examples of which are shown in Fig. 5. The intensity histograms associated with each timetrace reveal distributions with more than one peak, as is expected when timetraces feature alternations of two or more states, each with a distinct average emission intensity; in accordance with our lifetime measurements, we posit $X_0$ and $X^{1-}$ as two non-dim states with two *a priori* different emission quantum yields and therefore unequal brightness, $X^{1+}$ and XX as the dim/off state. (Such an "off" state can still contribute a noticeable number of signal photons to a decay curve if it occurs for a long-enough total time during the measurement.) Unfortunately, the exact shape of the intensity histogram is highly complex in such cases due to the presence of "mixed" bins with unresolved short transitions between states and the influence of shot noise [70, 71]; additional hypotheses about the nature of the underlying state transitions are necessary but not easily justifiable because decomposing intensity trace into a putative sequence of ON, DIM and OFF periods by means of a threshold criterion can lead to potentially biased and inconsistent results [71, 72]. As it is nevertheless important, in view of potential applications, to quantify the fraction of time that a given NC species spends in emissive states, we have adopted a simple protocol to assign an estimate (at a certain level of confidence/statistical significance) of this ON fraction to each timetrace. To this end, a Poisson distribution was fitted to the low-intensity region in each count histogram to identify an average off-state count rate, which in turn allowed identifying and counting all timetrace bins that were unlikely to be OFF at the $p = 0.05$ level. The ON-state fraction thus deduced is not meant to be quantitative in an absolute sense, but rather to serve as a heuristic *comparative* figure of merit, whose main advantage is that it is determined by an algorithm that, once defined, works without further human intervention and treats all timetraces equally to minimize potential bias in the comparison.

**Table 2:** Median ON-time fraction and median time to photobleaching deduced from timetraces (duration: 300 s) of $n$ individual OA- or halide-capped NCs on glass substrates (day 0 = deposition day).

| NC sample | | $n$ | ON fraction | bleaching time |
|---|---|---|---|---|
| OA-capped | | 21 | 81 % | $> 300$ s |
| halide-capped | day 0 | 12 | 90 % | 120 s |
| | day 1 | 24 | 64 % | $\geqslant 300$ s |
| | day 2 | 10 | 57 % | $\geqslant 300$ s |

The blinking statistics of individual OA-capped and halide-capped NCs deposited on glass substrates are summarized in Table 2; given that the PL spectra of the halide-capped variety showed pronounced aging effects within a few days in a film on glass, Fig. 2 c, we differentiated the blinking behavior on different days after deposition in this case. The OA-capped NCs exhibited a large median ON-time fraction and were found to be highly resistant to photobleaching, with 100 % of the studied NCs surviving the full observation period of five minutes. The halide-capped QDs showed an even higher ON-time fraction on deposition day combined with a significantly reduced photostability. Interestingly, the resistance to bleaching of the halide-capped NCs became comparable again to the OA-capped species from day 1 on, while a moderate reduction of the ON-time fraction was observed; as a matter of fact, the increase in photostability with time was already clearly noticeable over the course of the experiments on day 0. This is in line with the ensemble time-dependent intensity measurements performed over one week on halide-capped NC thin films, which show a decrease of the PL over the first day, to see a stabilization of the signal over the next days of the experiments (see Fig. S3 c of the ESI†).

To further quantify the differences in the blinking statistics of the four samples, we applied a Kolmogorov–Smirnov (K-S) test to the distributions of ON-time fractions found in each case, which yielded $p$-values for the null hypothesis that an observed difference between a given pair of distributions is caused by mere statistical fluctuations rather than a real change in the underlying photophysical parameters. If we then adopt the usual $p < 0.05$ criterion for statistical significance, we can draw the following conclusions: The ON-time difference between the OA-capped NCs and the halide-capped NCs measured right after deposition (day 0) was not significant ($p = 0.1$), while the discrepancy between the halide-capped NCs on day 0 versus days 1 and 2 was significant ($p = 0.01$ in both cases); the main aging process seemed to occur between day 0 and day 1, as the variation between days 1 and 2 was of minor significance once again ($p = 0.8$). It thus appears to be the case that the OA-to-halide ligand exchange initially does not modify the interaction potential between the surface and the confined exciton in ways that dramatically alter the blinking behavior. One possible explanation of this observation is that the thick (almost 6 nm) type-I gradient shell confines the exciton so efficiently as to minimize the interactions with the NC surface, which might indicate that blinking is not dominated by surface traps but rather by volume traps situated in the shell at shorter distances from the NC core [73]. Alternatively one might surmise that the surface passivation of halide-capped NCs is, at least initially, of comparable efficacy as the one provided by OA ligands, and that this passivation deteriorates subsequently due to the known tendency of halide ions to react with with ambient air and humidity [33]. The latter interpretation is supported by the fact that surface modification did seem to become relevant once the NCs had been exposed to the ambient atmosphere on glass slides for a certain time, consistent with the aging observed in the PL spectra of Fig. 2 c and the evolution of the PL intensity over one week shown in Fig. S3 c of the ESI†.

The blinking data thus allows us to pinpoint the major changes as occurring during an interval of several hours up to one day after deposition; once the aging process has run its course, the new blinking behavior remains stable for at least a couple of days. The encouraging conclusion that can be drawn from these findings with respect to technological applications of the halide-capped QDs is that both their emissivity (ON-time fraction) and their bleaching resistance does not deteriorate under air over periods of time that are sufficiently long to be compatible with procedures such as incorporation in protective matrices or photonic devices.

## 4 Conclusion

We have presented an experimental protocol for replacing the native oleic acid ligands of gradient-shell CdSe/CdZnS nanocrystals with halide ions ($Br^-$ and $Cl^-$). Ensemble absorption, photoluminescence excitation and photoluminescence spectra are unaffected by the ligand-exchange procedure and their evolution over time suggests that both the OA- and the halide-capped NCs are stable for several weeks in solution without significant degradation or modification of their photoluminescence emission. The halide-capped NCs are furthermore stable for at least a few hours when deposited on a bare glass slide and exposed to air; after 1 day, however, distinct spectral shifts and changes in the shape of the emission band do occur, which is well fitted by an increase of the size distribution. Individual NCs of both species could be studied on glass slides and are viable single-photon emitters, as demonstrated by photoluminescence antibunching measurements. Lifetime measurements point to the existence of four states with different emissivities, which we surmise to correspond to the neutral exciton ($X_0$), the $X^{1-}$ and $X^{1+}$ singly-charged excitons, and a mix of multi-charged single excitons and bi-excitons (XX), respectively. No significant change in the excited-state lifetimes of the is observed when switching from OA- to halide-capped NCs, but a transfer of emission probability from $X_0$ to the fastest decay components does take place, while $X^{1-}$ remains the dominant emissive states in both species. Blinking statistics of single NCs confirm that the pertinent photophysical properties of halide-capped NCs remain comparable to the OA-capped NCs even when the bare halide-capped NCs are exposed to ambient air on glass substrates for a few hours, which suggests that the halide ions can initially maintain a surprisingly efficient passivation of the NC surface. After one day of exposition to ambient conditions, statistically significant changes in the blinking statistics do occur, which moderately reduces the fraction of time spent in the bright state(s) under continuous excitation while increasing the resistance to photobleaching, which then becomes comparable to that of the precursor-NCs with the original OA ligands. These observations suggest that the aging process has a partially protective effect against photobleaching, which is also coherent with the time-evolution of the emission intensity of the halide-capped thin films. Our work shows that the remarkable optical properties of CdSe/CdZnS NCs emitting in the visible can be largely conserved when the OA-ligands are replaced by halide ions such as $Br^-$ and $Cl^-$, including maintaining a high-enough quantum yield for single-QD experiments to remain feasible. These encouraging results suggest that halide-capped NCs are promising candidates for active species in quantum-dot-based devices in cases where the presence of bulky organic ligands can be problematic, for example when incorporation into and/or efficient coupling with hybrid organic/inorganic matrices such as halide perovskites is required for, e. g., light-harvesting or single-photon-source applications. Furthermore, the documented shift from neutral-exciton to charged-exciton emission induced by the ligand exchange might be interesting for applications in photocatalysis.

**Conflicts of interest**

There are no conflicts to declare.

**Acknowledgements**

The authors thank J.-F. Sivignon, Y. Guillin, G. Montagne and the Lyon center for nano-opto technologies (NanOpTec) for technical support. This work was made possible by financial support from the IDEXLYON project of Université de Lyon in the framework of the Investissements d'Avenir program (ANR16-IDEX-0005), with further funding from the Agence Nationale de Recherche (ANR-16-CE24-0002) and the Fédération de Recherche André Marie Ampère (FRAMA).

**References**

[1] M. Y. Han, X. H. Gao, J. Z. Su and S. Nie, *Nat. Biotechnol.*, 2001, **19**, 631–635.

[2] S. Chang, M. Zhou and C. P. Grover, *Opt. Express*, 2004, **12**, 143–148.

[3] P. O. Anikeeva, J. E. Halpert, M. G. Bawendi and V. Bulović, *Nano Lett.*, 2009, **9**, 2532–2536.

[4] V. Wood and V. Bulović, *Nano Rev.*, 2010, **1**, 5202.

[5] D. V. Talapin, J. S. Lee, M. V. Kovalenko and E. V. Shevchenko, *Chem. Rev.*, 2010, **110**, 389–458.

[6] Y. Shirasaki, G. J. Supran, M. G. Bawendi and V. Bulović, *Nat. Photonics*, 2013, **7**, 13–23.

[7] J. Bao and M. G. Bawendi, *Nature*, 2015, **523**, 67–70.

[8] G. H. Carey, A. L. Abdelhady, Z. J. Ning, S. M. Thon, O. M. Bakr and E. H. Sargent, *Chem. Rev.*, 2015, **115**, 12732–12763.

[9] Y. S. Park, A. V. Malko, J. Vela, Y. Chen, Y. Ghosh, F. García-Santamaría, J. A. Hollingsworth, V. I. Klimov and H. Htoon, *Phys. Rev. Lett.*, 2011, **106**, 187401.

[10] B. G. Jeong, Y. S. Park, J. H. Chang, I. Cho, J. K. Kim, H. Kim, K. Char, J. Cho, V. I. Klimov, P. Park, D. C. Lee and W. K. Bae, *ACS Nano*, 2016, **10**, 9297–9305.

[11] O. Labeau, P. Tamarat and B. Lounis, *Phys. Rev. Lett.*, 2003, **90**, 257404.

[12] O. Labeau, P. Tamarat and B. Lounis, *Appl. Phys. Lett.*, 2006, **88**, 223110.

[13] C. B. Murray, D. J. Norris and M. G. Bawendi, *J. Am. Chem. Soc.*, 1993, **115**, 8706–8715.

[14] Z. J. Ning, X. W. Gong, R. Comin, G. Walters, F. J. Fan, O. Voznyy, E. Yassitepe, A. Buin, S. Hoogland and E. H. Sargent, *Nature*, 2015, **523**, 324–328.

[15] H. S. Kim, C. R. Lee, J. H. Im, K. B. Lee, T. Moehl, A. Marchioro, S. J. Moon, R. Humphry-Baker, J. H. Yum, J. E. Moser, M. Grätzel and N. G. Park, *Sci Rep*, 2012, **2**, 591.

[16] P. Gao, M. Grätzel and M. K. Nazeeruddin, *Energy Environ. Sci.*, 2014, **7**, 2448–2463.

[17] A. Swarnkar, R. Chulliyil, V. K. Ravi, M. Irfanullah, A. Chowdhury and A. Nag, *Angew. Chem.-Int. Edit.*, 2015, **54**, 15424–15428.

[18] N. G. Park, M. Grätzel, T. Miyasaka, K. Zhu and K. Emery, *Nat. Energy*, 2016, **1**, 16152.

[19] M. Nirmal, B. O. Dabbousi, M. G. Bawendi, J. J. Macklin, J. K. Trautman, T. D. Harris and L. E. Brus, *Nature*, 1996, **383**, 802–804.

[20] M. A. Boles, D. Ling, T. Hyeon and D. V. Talapin, *Nat. Mater.*, 2016, **15**, 141–153.

[21] M. S. Kodaimati, K. P. McClelland, C. He, S. C. Lian, Y. S. Jiang, Z. Y. Zhang and E. A. Weiss, *Inorg. Chem.*, 2018, **57**, 3659–3670.

[22] R. Burke, K. L. Bren and T. D. Krauss, *J. Chem. Phys.*, 2021, **154**, 030901.

[23] X. G. Peng, M. C. Schlamp, A. V. Kadavanich and A. P. Alivisatos, *J. Am. Chem. Soc.*, 1997, **119**, 7019–7029.

[24] S. Jeong, M. Achermann, J. Nanda, S. Lvanov, V. I. Klimov and J. A. Hollingsworth, *J. Am. Chem. Soc.*, 2005, **127**, 10126–10127.

[25] J. W. Stouwdam, J. Shan, F. C. J. M. van Veggel, A. G. Pattantyus-Abraham, J. F. Young and M. Raudsepp, *J. Phys. Chem. C*, 2007, **111**, 1086–1092.

[26] Y. Kim, N. W. Song, H. Yu, D. W. Moon, S. J. Lim, W. Kim, H. J. Yoon and S. K. Shin, *Phys. Chem. Chem. Phys.*, 2009, **11**, 3497–3502.

[27] A. D. Zhang, C. Q. Dong, H. Liu and J. C. Ren, *J. Phys. Chem. C*, 2013, **117**, 24592–24600.

[28] X. Y. Li, Y. B. Zhao, F. J. Fan, L. Levina, M. Liu, R. Quintero-Bermudez, X. W. Gong, L. N. Quan, J. Z. Fan, Z. Y. Yang, S. Hoogland, O. Voznyy, Z. H. Lu and E. H. Sargent, *Nat. Photonics*, 2018, **12**, 159.

[29] J. J. Buckley, M. J. Greaney and R. L. Brutchey, *Chem. Mat.*, 2014, **26**, 6311–6317.

[30] D. N. Dirin, S. Dreyfuss, M. I. Bodnarchuk, G. Nedelcu, P. Papagiorgis, G. Itskos and M. V. Kovalenko, *J. Am. Chem. Soc.*, 2014, **136**, 6550–6553.

[31] R. M. Dragoman, M. Grogg, M. I. Bodnarchuk, P. Tiefenboeck, D. Hilvert, D. N. Dirin and M. V. Kovalenko, *Chem. Mat.*, 2017, **29**, 9416–9428.

[32] D. Bederak, D. M. Balazs, N. V. Sukharevska, A. G. Shulga, M. Abdu-Aguye, D. N. Dirin, M. V. Kovalenko and M. A. Loi, *ACS Appl. Nano Mater.*, 2018, **1**, 6882–6889.

[33] F. Purcell-Milton, M. Chiffoleau and Y. K. Gun'ko, *J. Phys. Chem. C*, 2018, **122**, 25075–25084.

[34] W. R. Wang, Z. X. Pan, H. S. Rao, G. Z. Zhang, H. Song, Z. Y. Zhang and X. H. Zhong, *Chem. Mat.*, 2020, **32**, 630–637.

[35] B. Fisher, J. M. Caruge, D. Zehnder and M. Bawendi, *Phys. Rev. Lett.*, 2005, **94**, 087403.

[36] O. Carion, B. Mahler, T. Pons and B. Dubertret, *Nat. Protoc.*, 2007, **2**, 2383–2390.

[37] P. T. Jing, J. J. Zheng, Q. H. Zeng, Y. L. Zhang, X. M. Liu, X. Y. Liu, X. G. Kong and J. L. Zhao, *J. Appl. Phys.*, 2009, **105**, 044313.

[38] L. Li, G. J. Tian, Y. Luo, H. Brismar and Y. Fu, *J. Phys. Chem. C*, 2013, **117**, 4844–4851.

[39] B. Mahler, P. Spinicelli, S. Buil, X. Quélin, J. P. Hermier and B. Dubertret, *Nat. Mater.*, 2008, **7**, 659–664.

[40] W. Qin, R. A. Shah and P. Guyot-Sionnest, *ACS Nano*, 2012, **6**, 912–918.

[41] J. Lim, B. G. Jeong, M. Park, J. K. Kim, J. M. Pietryga, Y. S. Park, V. I. Klimov, C. Lee, D. C. Lee and W. K. Bae, *Adv. Mater.*, 2014, **26**, 8034–8040.

[42] S. Y. Park, H. S. Kim, J. Yoo, S. Kwon, T. J. Shin, K. Kim, S. Jeong and Y. S. Seo, *Nanotechnology*, 2015, **26**, 275602.

[43] B. M. Koo, S. Sung, C. X. Wu, J. W. Song and T. W. Kim, *Sci Rep*, 2019, **9**, 9755.

[44] H. Xu, H. Brismar and Y. Fu, *J. Colloid Interface Sci.*, 2017, **505**, 528–536.

[45] Y. Fu, J. Jussi, L. Elmlund, S. Dunne, Q. Wang and H. Brismar, *Phys. Rev. B*, 2019, **99**, 035404.

[46] A. M. Munro, I. Jen-La Plante, M. S. Ng and D. S. Ginger, *J. Phys. Chem. C*, 2007, **111**, 6220–6227.

[47] D. Magde, J. H. Brannon, T. L. Cremers and J. Olmsted, *J. Phys. Chem.*, 1979, **83**, 696–699.

[48] A. Aubret, J. Houel, A. Pereira, J. Baronnier, E. Lhuillier, B. Dubertret, C. Dujardin, F. Kulzer and A. Pillonnet, *ACS Appl. Mater. Interfaces*, 2016, **8**, 22361–22368.

[49] Z. Bajzer, T. M. Therneau, J. C. Sharp and F. G. Prendergast, *Eur. Biophys. J. Biophys. Lett.*, 1991, **20**, 247–262.

[50] M. Köllner and J. Wolfrum, *Chem. Phys. Lett.*, 1992, **200**, 199–204.

[51] B. Lounis, H. A. Bechtel, D. Gerion, P. Alivisatos and W. E. Moerner, *Chem. Phys. Lett.*, 2000, **329**, 399–404.

[52] W. H. Press, S. A. Teukolsky, W. T. Vetterling and B. P. Flannery, *Numerical Recipes: The Art of Scientific Computing*, Cambridge University Press, Cambridge, 3rd edn., 2007.

[53] L. E. Brus, *J. Chem. Phys.*, 1984, **80**, 4403–4409.

[54] M. S. Kang, A. Sahu, D. J. Norris and C. D. Frisbie, *Nano Lett.*, 2010, **10**, 3727–3732.

[55] C. Javaux, B. Mahler, B. Dubertret, A. Shabaev, A. V. Rodina, A. L. Efros, D. R. Yakovlev, F. Liu, M. Bayer, G. Camps, L. Biadala, S. Buil, X. Quélin and J. P. Hermier, *Nat. Nanotechnol.*, 2013, **8**, 206–212.

[56] D. Canneson, L. Biadala, S. Buil, X. Quélin, C. Javaux, B. Dubertret and J. P. Hermier, *Phys. Rev. B*, 2014, **89**, 035303.

[57] X. Hou, H. Qin and X. Peng, *Nano Lett.*, 2021, **21**, 3871–3878.

[58] P. Spinicelli, S. Buil, X. Quélin, B. Mahler, B. Dubertret and J. P. Hermier, *Phys. Rev. Lett.*, 2009, **102**, 136801.

[59] O. Chen, J. Zhao, V. P. Chauhan, J. Cui, C. Wong, D. K. Harris, H. Wei, H. S. Han, D. Fukumura, R. K. Jain and M. G. Bawendi, *Nat. Mater.*, 2013, **12**, 445–451.

[60] X. Brokmann, G. Messin, P. Desbiolles, E. Giacobino, M. Dahan and J. P. Hermier, *New J. Phys.*, 2004, **6**, 99.

[61] L. W. Wang, M. Califano, A. Zunger and A. Franceschetti, *Phys. Rev. Lett.*, 2003, **91**, 056404.

[62] P. P. Jha and P. Guyot-Sionnest, *ACS Nano*, 2009, **3**, 1011–1015.

[63] C. Galland, Y. Ghosh, A. Steinbrück, J. A. Hollingsworth, H. Htoon and V. I. Klimov, *Nat. Commun.*, 2012, **3**, 908.

[64] F. García-Santamaría, S. Brovelli, R. Viswanatha, J. A. Hollingsworth, H. Htoon, S. A. Crooker and V. I. Klimov, *Nano Lett.*, 2011, **11**, 687–693.

[65] W. K. Bae, L. A. Padilha, Y. S. Park, H. McDaniel, I. Robel, J. M. Pietryga and V. I. Klimov, *ACS Nano*, 2013, **7**, 3411–3419.

[66] D. I. Chepic, A. L. Efros, A. I. Ekimov, M. G. Vanov, V. A. Kharchenko, I. A. Kudriavtsev and T. V. Yazeva, *J. Lumines.*, 1990, **47**, 113–127.

[67] M. Kuno, D. P. Fromm, H. F. Hamann, A. Gallagher and D. J. Nesbitt, *J. Chem. Phys.*, 2000, **112**, 3117–3120.

[68] R. Verberk, J. W. M. Chon, M. Gu and M. Orrit, *Physica E*, 2005, **26**, 19–23.

[69] A. A. Cordones and S. R. Leone, *Chem. Soc. Rev.*, 2013, **42**, 3209–3221.

[70] R. T. Sibatov and V. V. Uchaikin, *Opt. Spectrosc.*, 2010, **108**, 761–767.

[71] N. Amecke, A. Heber and F. Cichos, *J. Chem. Phys.*, 2014, **140**, 114306.

[72] C. H. Crouch, O. Sauter, X. H. Wu, R. Purcell, C. Querner, M. Drndic and M. Pelton, *Nano Lett.*, 2010, **10**, 1692–1698.

[73] R. Verberk, A. M. van Oijen and M. Orrit, *Phys. Rev. B*, 2002, **66**, 233202.